\documentclass[useAMS,usenatbib]{mn2e}

\usepackage{amssymb}
\usepackage{amsfonts}
\usepackage{mncite}
\usepackage{epsfig}
\usepackage{psfig}

\newcommand{\de}{\mbox{d}}

\title[Accelerated planetesimal growth in self-gravitating
      discs] {Accelerated planetesimal growth in
      self-gravitating protoplanetary discs}

\author[Rice et al.]{W.K.M. Rice$^1$, G. Lodato$^2$,
        J.E. Pringle$^{2,3}$, P.J. Armitage$^{4,5}$ and
        I.A. Bonnell$^1$ \\ $^1$School of Physics and Astronomy,
        University of St Andrews, North Haugh, St Andrews KY16 9SS \\
        $^2$Institute of Astronomy, Madingley Road, Cambridge, CB3 0HA
        \\ $^3$Space Telescope Science Institute, 3700 San Martin
        Drive, Baltimore, MD 21218, USA \\ $^4$JILA, Campus Box 440,
        University of Colorado, Boulder CO 80309-0440, USA \\
        $^5$Department of Astrophysical and Planetary Sciences,
        University of Colorado, Boulder CO 80309-0391, USA }
\begin{document}

\maketitle

\begin{abstract}
In this paper we consider the evolution of small planetesimals 
(radii $\sim 1 - 10$ metres) in
marginally stable, self-gravitating protoplanetary discs. The drag
force between the disc gas and the embedded planetesimals generally
causes the planetesimals to drift inwards through the disc at a rate
that depends on the particle size. In a marginally stable,
self-gravitating disc, however, the planetesimals are significantly
influenced by the non-axisymmetric spiral structures resulting from
the growth of the gravitational instability. The drag force now causes
the planetesimals to drift towards the peaks of the spiral arms where
the density and pressure are highest. For small particles, that are
strongly coupled to the disc gas, and for large particles, that have
essentially decoupled from the disc gas, the effect is not
particularly significant.  Intermediate sized particles, which would
generally have the largest radial drift rates, do, however, become
significantly concentrated at the peaks of the spiral arms. These high
density regions may persist for, of order, an orbital period and may
attain densities comparable to that of the disc gas. Although at the
end of the simulation only $\sim 25$ \% of the planetesimal particles
lie in regions of enhanced density, during the course of the
simulation at least $75$ \% of the planetesimal particles have at some
stage been in a such a region. We find that the concentration of
particles in the spiral arms results in an increased collision rate,
an effect that could significantly accelerate planetesimal growth. The
density enhancements may also be sufficient for the growth of
planetesimals through direct gravitational collapse. The interaction
between small planetesimals and self-gravitating spiral structures may
therefore play an important role in the formation of large
planetesimals that will ultimately coagulate to form terrestrial
planets or the cores of gas/ice giant planets.
\end{abstract}


\section{Introduction}
The formation of both terrestrial planets and the cores of gas/ice
giant planets is thought to occur through the collisional accumulation
of planetesimals \citep{safronov,wetherill90,weidencuzzi93}. In the
case of gas giant planets, a gaseous envelope is accreted once the
core has become sufficiently massive \citep{lissauer93,pollack96}.
Since this must occur while there is still enough gas in the
circumstellar disc, and since observations \citep{haisch01} suggest
that most stars lose their gaseous discs within $\sim 10^7$ years, it is
generally accepted that gas giant planets have to form within $\sim 10^7$
years.

The standard core accretion models of giant planet formation
\citep{pollack96,bodenheimer2000} suggest formation times that could
easily exceed disc lifetimes. However, this long formation timescale
problem might be solved in light of the results of recent numerical
simulations of the evolution of planetary cores embedded in turbulent
accretion discs \citep{nelson2004}. These simulations show that cores
may actually undergo a random walk through the disc, leading to the
suggestion that core migration may significantly accelerate gas giant
formation \citep{ricearmi03,alibert04}, a possibility initially
recognised by \citet{hourigan84}.

An additional difficulty in the standard core accretion model is the
growth of kilometre sized planetesimals from, initially, micron sized
dust grains. Within a few scaleheights of the disc midplane, the
pressure gradient in the circumstellar disc tends, for standard disc
geometries, to be negative and causes the gas to orbit with
sub-Keplerian velocities. The dust, which is not affected by the gas
pressure gradient, and the gas therefore orbit with different
velocities and the resulting drag force causes the dust grains 
to drift inwards at
a rate that depends on their size \citep{weiden77,tak02}.  For
small sizes, the dust grains are essentially coupled to the disc gas
and the radial drift velocity is consequently small. For large sizes,
the grains are decoupled from the gas, move in nearly Keplerian
orbits, and again have small radial drift velocities. Particles with
intermediate sizes can, however, have large inward radial
velocities. Although the exact size range depends on the circumstellar
disc properties, the maximum radial velocity may easily exceed
$10^3\mbox{cm/s}$ and is normally thought to occur for objects with
sizes between 1cm and 1m \citep{weiden77}.

Although the differential radial velocity is a crucial part of the
grain growth, since larger objects grow by sweeping up smaller objects
(e.g., \citealt{safronov,weidencuzzi93}), if the maximum radial
velocity is too high these objects may drift inwards before they can
become large enough to decouple from the disc gas. Together with the
inward radial drift, the larger grains may also settle towards the
midplane \citep{goldreichward,garaud04}, producing a thin, dense dust
layer (note, however, that the presence of turbulence in the disc can
prevent the dust settling, see \citealt{stone96}). This layer may
become self-gravitating and, if sufficiently unstable, could produce
kilometre sized objects directly via gravitational collapse
\citep{goldreichward}. This, however, requires extremely small random
dust velocities ($\sim 10$ cm/s), which may be difficult to achieve
\citep{weidencuzzi93,cuzzi93}. Even if such small random velocities
are possible, if objects are to grow via gravitational collapse, any
increase in the random velocities has to be lost very rapidly
\citep{gammie01,rice03a}.

An alternative mechanism for the formation of gas giant planets, that
does not require the initial growth of a core, is that the gas itself
may become unstable, producing gravitationally bound gaseous
protoplanets \citep{boss98,boss00}.  This again requires that the gas
be cold and that it remain `almost isothermal'
\citep{pickett98,pickett2000} or, equivalently, that any cooling
mechanism is extremely efficient
\citep{gammie01,rice03a,gammie03}. Furthermore, some simulations of
fragmenting protoplanetary discs suggest that, at best, this mechanism
may be able to produce only the most massive ($> 5$ Jupiter masses)
gas giant planets \citep{rice03b}.

We study here the influence of the development of gravitationally
unstable spiral modes on the planet formation mechanism via core
accretion \citep[see also][]{haghighipour03a}.  Although it is
possible that the conditions required for disc fragmentation may never
be met \citep{pickett2000}, it is quite likely that protostellar discs
are self-gravitating in their early stages
\citep[e.g.,][]{lin90,LB2001}. If so, the disc then becomes
susceptible to the growth of non-axisymmetric spiral structures which
can transport angular momentum very efficiently
\citep[e.g.,][]{LR04}. In the presence of such spiral structures, the
gas pressure gradient, which can be large, changes from positive on
one side of the structure to negative on the other, resulting in both
super- and sub-Keplerian gas velocities. The gas drag force then
causes dust grains to drift both radially inwards and outwards,
depending on whether the local gas velocity is super-Keplerian or
sub-Keplerian \citep{haghighipour03a,haghighipour03b}. The net effect
is that the dust drifts towards the density maxima, where the pressure
gradient is zero \citep{haghighipour03a}. A similar effect would occur
in the presence of any coherent and long-lived density enhancement. 
Previous work
\citep{godon00, klahr03} has considered how vorticies may influence
embedded particles. The formation of such vorticies is, however, still
a matter of some debate.

In this paper we present the results of three-dimensional, global
simulations of self-gravitating accretion discs in which we include
both gas and dust, coupled via a drag force
\citep{whipple72,weiden77}. The gaseous disc is maintained in a state
of marginal gravitational instability, achieved by letting the disc
cool down (through a simple parametrisation of the cooling function;
\citealt{gammie01} and \citealt{rice03a}).  In this way a quasi-steady
spiral structure develops in the disc and is maintained during several
dynamical time-scales. We are therefore able to follow the process of
concentration of the planetesimals in the spiral arms. We have
performed several simulations, considering planetesimals of different
sizes. We find that, for a given size range, the planetesimals are
indeed able to reach high concentrations, in some regions attaining
densities comparable to the gas. This could significantly enhance the
coagulation of planetesimals into larger bodies, by increasing the
planetesimal collision rates and/or by making the planetesimal
sub-disc become gravitationally unstable
\citep{youdinshu02,youdinchiang04}.

This paper is organised as follows. In Section \ref{sec:dynamics} we
summarize the basic features of the coupled dynamics of a two
component (gas and planetesimals) disc, including a description of the
relevant drag forces. In Section \ref{sec:numerical} we describe the
simulation code that we have used and the numerical setup. In Section
\ref{sec:results} we describe our results. In Section
\ref{sec:conclusion} we draw our conclusions.
    
\section{Gas-planetesimal disc dynamics}

\label{sec:dynamics}

We consider a system comprising two interpenetrating discs: a ``gas''
disc, that is evolved using the standard hydrodynamical equations of
motion, and a ``planetesimal'' disc, that is considered as a
collection of test particles evolved under the influence of
gravitational and drag forces alone. Both the ``gas'' disc and the
``planetesimal'' disc rotate around a central protostar of mass
$M_{\star}$, which we take to be equal to $1M_{\odot}$. The term
``planetesimal'' is generally used to refer to particles that have
decoupled from the disc gas.  Although the particles we consider here
are still coupled to the gas, we use this term because, as we will
discuss later, the particles we consider are clearly too large to be
regarded as dust grains.

In order to illustrate the basic dynamical ingredients of our model,
and to introduce the relevant physical quantities, let us consider
first the simple case of a smooth, axisymmetric, non self-gravitating
disc. We define the Keplerian velocity, $V_{\mathrm K}$, around the
star at a distance $R$ by
\begin{equation}
\label{eq:keplerian}
V_{\mathrm K}^2=GM_{\star}/R.
\end{equation} 
The gas disc is characterized by a surface density profile
$\Sigma(R)$, a temperature profile $T(R)$ and a pressure profile
$P(R)$. The gas sound speed is defined as
\begin{equation}
\label{eq:cs}
c_{\mathrm s}^2=\frac{\partial P}{\partial \rho},
\end{equation}
where $\rho$ is the gas volume density. Let ${\mathbf v}$ and
 ${\mathbf v}_{\mathrm{p}}$ be the velocities of the gas and of the
 planetesimals, respectively. In centrifugal equilibrium the azimuthal
 component of the gas velocity $v_{\phi}$ is given by
\begin{equation}
\label{eq:gasvel}
v_{\phi}^2=V_K^2+\left(\frac{\de\ln\rho}{\de\ln R}\right)
c_{\mathrm s}^2=V_K^2(1+\gamma),
\end{equation}
where $\gamma\equiv(\de\ln\rho/\de\ln R)c_{\mathrm s}^2/V_K^2$, is a
measure of the importance of thermal effects in the disc. For most
disc models, the density $\rho$ decreases with radius, so that
$\gamma<0$ and $v_{\phi}$ is generally sub-Keplerian. If the gas disc
is in vertical hydrostatic equilibrium, its aspect ratio is given by
\begin{equation}
\frac{H}{R}\approx\frac{c_{\mathrm
s}}{V_K}=\left|\frac{\de\ln\rho}{\de\ln R}\right|^{-1/2} \sqrt{|\gamma|}
\end{equation}

The planetesimals, on the other hand, are not affected by pressure
forces and, in centrifugal equilibrium therefore orbit at the
Keplerian velocity $V_K$. Let ${\mathbf u}$ be the relative velocity
between the gas and the planetesimals. We assume that the gas and the
planetesimals are coupled through a drag force, as described by
\citet{weiden77} and \citet{whipple72}. The drag force is given by
\begin{equation}
{\mathbf F}_{\mathrm{D}}=-\frac{1}{2}C_{\mathrm{D}}\pi
a^2\rho u^2\hat{\mathbf u},
\end{equation}
where $u=|{\mathbf u}|$, $\hat{\mathbf u}={\mathbf u}/u$, $a$ is the
mean radius of the planetesimals and $C_{\mathrm{D}}$ is the drag
coefficient, given by
\begin{equation}
C_{\mathrm{D}} = \left\{ \begin{array}{ccc}
        \displaystyle \frac{8}{3}\frac{c_s}{u} & ~~~~~~~~ & a<9\lambda/4\\ 
\\
        24R_e^{-1}                              & ~~~~~~~~ & R_e<1 \\
\\
        24R_e^{-0.6}                            & ~~~~~~~~ & 1<R_e<800\\
\\
        0.44                                    & ~~~~~~~~ & R_e > 800
                 \end{array} \right. 
\end{equation}
The drag regime where $a<9\lambda/4$ is generally called the Epstein
regime, whereas the other three regimes define the Stokes drag.  In
the previous expression, $R_e$ is the Reynolds number, defined below
in Eq. (\ref{eq:reynold}) and $\lambda$ is the mean free path of the
gas particles. Assuming the gas to be made mainly of molecular
hydrogen, this is given by
\begin{equation}
\lambda=\frac{m_{H_2}}{\rho A}\approx \frac{4~10^{-9}}{\rho}~ \mbox{cm},
\end{equation}
where in the last expression the gas density $\rho$ is evaluated in
cgs units, $m_{H_2}$ is the mass of the hydrogen molecule, and $A$ is
its cross section
\begin{equation}
A=\pi a_0^2\approx 7~10^{-16} \mbox{cm}^2,
\end{equation}
where $a_0$ is the mean radius of the hydrogen molecule (see
\citealt{supulver00}).

The Reynolds number $R_e$ is given by
\begin{equation}
\label{eq:reynold}
R_e=\frac{2a\rho u}{\eta},
\end{equation}
where $\eta=\rho\nu$ is the gas viscosity. For collisional viscosity,
we have
\begin{equation}
\eta=\frac{\rho c_s\lambda}{2}.
\end{equation}
We can therefore write
\begin{equation}
R_e=4\left(\frac{a}{\lambda}\right)\left(\frac{u}{c_s}\right).
\end{equation}
For the disc properties and particle sizes that we will consider here, the 
Mach number rarely exceeds unity and the Reynolds number generally falls in
the range $1 <R_e<800$. Only in the inner regions of the disc, where
${\mathbf u}$ is largest, does the Mach number exceed unity and the Reynolds
number exceed $800$.

Another important quantity is the ``stopping time''
$t_{\mathrm{e}}$, where
\begin{equation}
t_{\mathrm{e}}=\frac{m_{\mathrm{p}}u}{|F_{\mathrm{D}}|}.
\end{equation}
Here $m_{\mathrm{p}}=4\pi\rho_{\mathrm{s}}a^3/3$, is the mass of the
planetesimals and $\rho_{\mathrm{s}}$ is the internal density of the
planetesimals (Throughout the paper, we take $\rho_{\mathrm{s}}=3$
g/cm$^3$). With the previous definitions, we have
\begin{equation}
\frac{t_{\mathrm{e}}}{t_{\mathrm{d}}}=\frac{8}{3C_{\mathrm{D}}}
\left(\frac{\rho_{\mathrm{s}}}{\rho}\right)
\left(\frac{V_{\mathrm{K}}}{u}\right)
\left(\frac{a}{R}\right),
\end{equation}
where $t_{\mathrm{d}}=\Omega_{\mathrm{K}}^{-1}=R/V_{\mathrm{K}}$ is
the local dynamical timescale. The smaller
$t_{\mathrm{e}}/t_{\mathrm{d}}$, the more strongly the planetesimals
are coupled to the gas. It is well known \citep{weiden77} that the
gas-planetesimal interaction through the drag force causes the
planetesimals to migrate in the direction of increasing pressure
(i.e., inward, for a smooth, axisymmetric disc). The migration rate
depends on the particle size and there is
a range of particles sizes (the extent of which depends
on the gas disc properties) that are characterized by a relatively
large radial drift.

Consider, as an illustration, the average structure of one of the gas
discs obtained from the numerical simulations of self-gravitating
discs by \citet{LR04}. On average, these discs are characterized by a
power-law surface density profile, $\Sigma\propto R^{-1}$, and by an
approximately flat profile, with a value of order unity, of the
axisymmetric gravitational stability parameter $Q$ \citep{toomre64}
\begin{equation}
\label{eq:Q}
Q=\frac{c_{\mathrm{s}}\kappa}{\pi G\Sigma}.
\end{equation}
In the previous expression $\kappa$ is the epicyclic frequency, that,
for the nearly Keplerian discs considered here, is roughly equal to
the orbital frequency $\Omega$, to order $H^2/R^2$. The disc extends
from $R_{\mathrm{in}}=0.25$ au to $R_{\mathrm{out}}=25$ au. The total
disc mass is $M_{\mathrm{disc}}=0.25M_{\odot}$. The profile of $Q$
essentially determines the pressure structure in the disc. We have
then that $c_{\mathrm{s}}\propto R^{1/2}$, $H/R\propto R$, and
$\rho\propto R^{-3}$.

\begin{figure*}
\centerline{\epsfig{figure=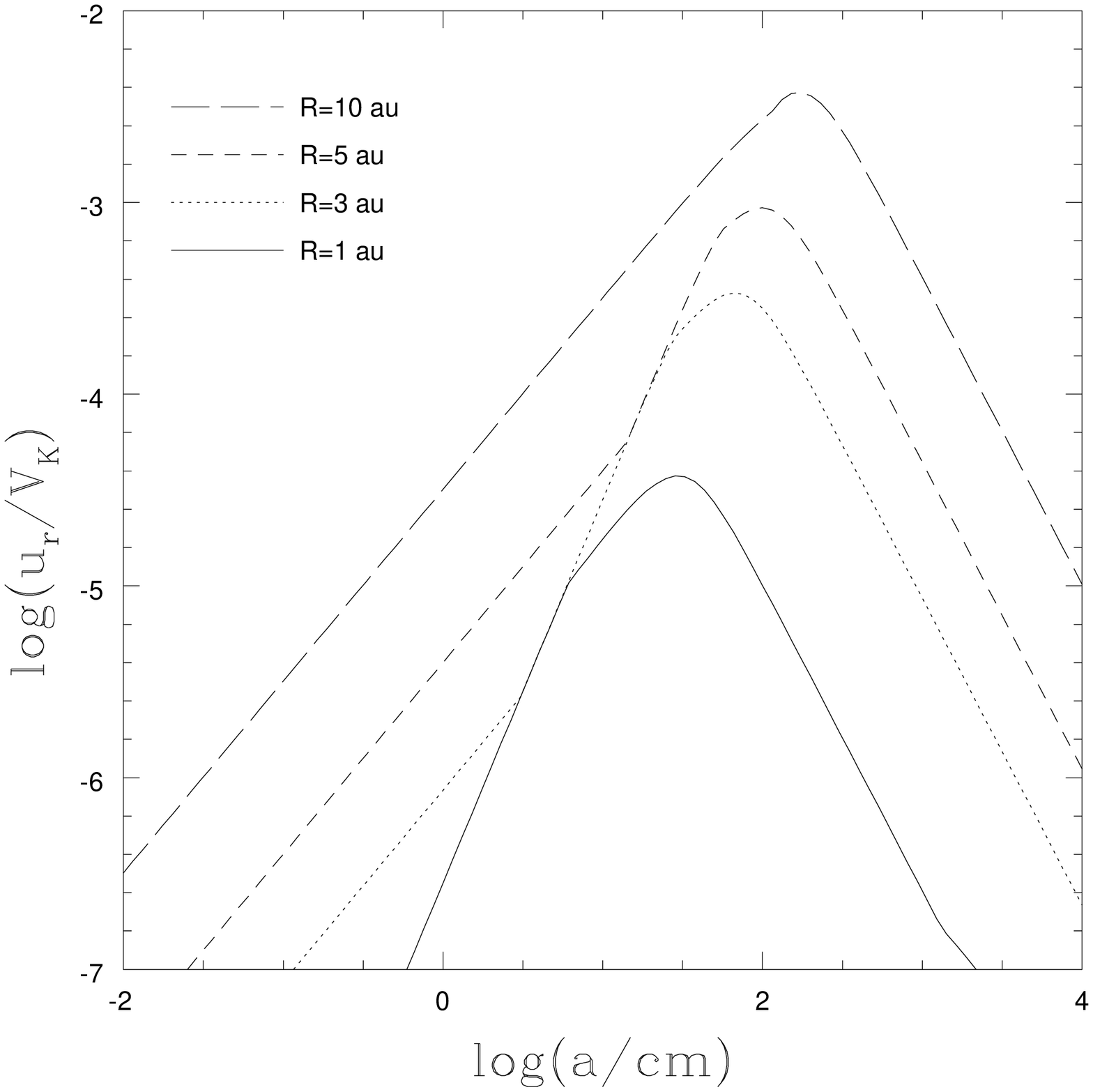,width=80mm}
            \epsfig{figure=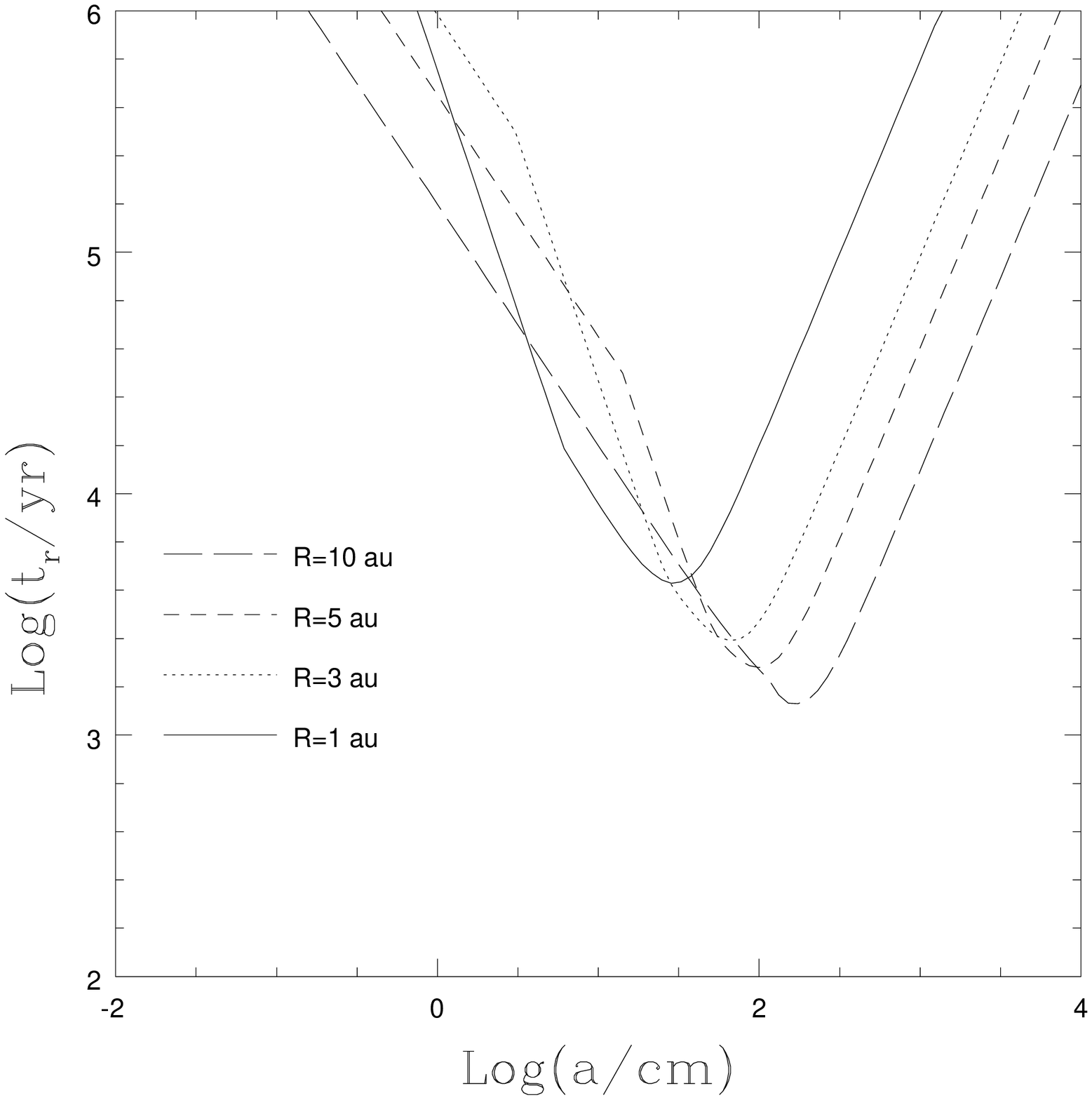,width=80mm}}
\caption{{\bf Left panel:} Radial drift velocity (in units of the
local Keplerian velocity) as a function of planetesimal size for the
assumed disc structure. {\bf Right panel:} Radial drift timescale as a
function of planetesimal size. The curves refer to four different
locations in the disc: (solid line) 1 au, (dotted line) 3 au, (dashed
line) 5 au, (long-dashed line) 10 au.}
\label{fig:drift}
\end{figure*} 

We can therefore estimate the radial drift velocity of planetesimals
in such a disc, using standard techniques \citep{weiden77}. The
results are shown in Fig. \ref{fig:drift}, where the left panel shows
the radial drift velocity $u_{\mathrm{r}}$ in units of the local
Keplerian velocity. Equivalently, defining the drift timescale
$t_{\mathrm{r}}= R/u_{\mathrm{r}}$, we have
$u_{\mathrm{r}}/V_{\mathrm{K}}= t_{\mathrm{d}}/ t_{\mathrm{r}}$, so
that the left panel of Fig. \ref{fig:drift} also shows a measure of
the drift timescale compared to the dynamical timescale. In the right
panel of Fig. \ref{fig:drift} we also show $t_{\mathrm{r}}$ in yrs. It
can be seen that for the assumed average gas disc properties, the
largest radial drift would occur for planetesimals with sizes in the
range between 10 and 100 cm. It should however be kept in mind that
these results refer to the azimuthally averaged structure of the
disc. Actually, the self-gravitating gas disc is characterized by a
spiral structure, with alternating regions of high and low gas
densities (see Fig. \ref{fig:gasspiral} below). Equation
(\ref{eq:gasvel}) then readily shows that the gas velocity changes
from sub-Keplerian through super-Keplerian, when moving across one arm
of the spiral structure. The drag force causes the planetesimals to
drift towards regions of higher pressure (and hence of higher
density), so we expect that the planetesimals tend to concentrate very
fast around the maxima of the gas density (see also
\citealt{haghighipour03a}). The results described above then suggest,
for the disc we are considering here, that this effect is maximal for
planetesimals with sizes between 10 and 100 cm. 

The expected drift timescale, in such structured discs, is going to be
smaller than in a smooth disc. In fact, in a smooth, power-law disc
the pressure gradient $\nabla P\sim P/R$, while in a disc with a
spiral structure (characterized by a typical scale $\sim H$) the
pressure gradient is of order $\nabla P\sim P/H$. The increased
pressure gradient decreases the drift timescale by a factor $H/R$. The
drift timescale is decreased by further factor $H/R$ because the
planetesimals have to move only a distance of order $H$ to reach the
density maxima. We therefore expect a significant concentration of the
planetesimals around the gas density maxima to take place over a
timescale a factor $(H/R)^2\sim 0.01$ smaller than that displayed in
Fig. \ref{fig:drift}, becoming therefore comparable with the dynamical
timescale in the most favourable cases. By inspection of the right
panel of Fig. \ref{fig:drift}, we also obtain that for particles
with sizes between $\sim 10$ and $\sim 1000$ cm, we need only run
our simulation for about $100$ yrs in order to follow
the effect of radial drift on the planetesimal sub-disc.

\section{Numerical simulations}

\label{sec:numerical}

\subsection{Smoothed particle hydrodynamics}

\label{sec:sph}

The three-dimensional gaseous disc used in these simulations is
modelled using smoothed particle hydrodynamics (SPH), a Lagrangian
hydrodynamics code (see \citealt{benz90,monaghan92}). The gas disc
consists of 250,000 SPH particles, each of which has a mass and an
internal energy (temperature). Each particle also has a smoothing
length that is allowed to vary with time to ensure that the number of
neighbours (SPH particles within 2 smoothing lengths) remains $\sim
50$. These neighbouring particles are used to determine the density
which, together with the internal energy, is used to compute the
pressure. The central star is modelled as a point mass onto which gas
particles may accrete if they approach to within the sink radius
(e.g., \citealt{bate95}), here taken to be 0.25 au. Both the central
point mass and the gas particles use a tree to determine gravitational
forces, and to determine the gas particle neighbours.

The small planetesimals are modelled using an additional type of
particle. These particles are, as far as the gas simulation is
concerned, massless (i.e., in these simulations we neglect the
self-gravity of the planetesimal disc and the back reaction of the
drag force on the gas). They experience only gravitational forces
(from the central star and from the disc gas) and are coupled to the
disc gas via a drag force. As discussed in Section \ref{sec:dynamics}
the drag force depends on the particle size, the local gas density,
and on the local gas velocity. To determine the drag force coefficient
also requires the gas sound speed which is calculated using the local
gas internal energy. The gravitational force on these test particles
is computed by including them in the tree. This also determines their
nearest gas neighbours. To ensure that the number of gas neighbours
remains $\sim 50$, the test particles also have a smoothing length
that is allowed to vary with time. These neighbouring particles are
then used to calculate the gas density, velocity, and internal energy
at the location of each test particle using the standard SPH
formalism (see \citealt{monaghan92}).  The exact value of the drag
force is then determined by specifying the planetesimal size. In each
simulation performed here, we use 125,000 test particles to 
represent a planetesimal disc which we assume contains 
particles of a single size.

An additional saving in computational time is made by using individual
particle time-steps \citep{bate95} with the time-steps for each
particle limited by the Courant condition and by a force condition
\citep{monaghan92}. As will be discussed in more detail in the
following sections, the gas disc is assumed, in the absence of any
cooling mechanism, to have an adiabatic equation of state with an
adiabatic index of $\gamma = 5/3$. The gas disc is initially evolved,
in the absence of test particles, by imposing a cooling term which is
chosen such that the disc ultimately settles into a quasi-steady,
self-gravitating state. The test particles are then added and the
simulation is evolved for an additional outer rotation period.
 
\subsection{Self-gravitating gas disc simulations}

\label{sec:selfgravity}

In this work we use the results of the numerical simulations by
\citet{LR04} of the dynamics of self-gravitating gas discs as an input
for our two-component gas-planetesimals disc simulations.

It is well known that the onset of gravitational instabilities in the
disc is determined by the value of the parameter $Q$, defined in
Eq. (\ref{eq:Q}). If $Q$ is smaller than a threshold value of order
unity, the disc quickly develops a spiral structure on the dynamical
timescale. The presence of the spiral influences strongly the thermal
evolution of the disc, in that it provides a source of effective
heating. Since $Q$ is proportional to the thermal speed
$c_{\mathrm{s}}$, in the absence of any cooling $Q$ would rapidly
become relatively large and the spiral structure would
vanish. However, if some cooling is present, a self-regulated state
can be achieved where the heating provided by the spiral structure
balances the external cooling, leading to a long-lasting spiral.

These processes have been recently explored numerically by
\citet{LR04} (see also \citealt{gammie01}), who have performed global,
three-dimensional simulations of the evolution of self-gravitating
discs. They imposed a cooling of the form:

\begin{equation}
\label{eq:cooling}
\left.\frac{\de U}{\de t}\right|_{\mathrm{cool}}=
-\frac{U}{t_{\mathrm{cool}}},
\end{equation}
where $U$ is the internal energy of the gas and the cooling timescale
is taken to be simply proportional to the dynamical timescale,
$t_{\mathrm{cool}}=\beta\Omega^{-1}$. For very rapid cooling
timescales ($\beta\lesssim 3$) the self-gravitating disc undergoes
fragmentation \citep{gammie01,rice03a}. In this work we have taken
$\beta=7.5$, giving a cooling time that should not, and indeed does
not, lead to fragmention.

\begin{figure}
\centerline{\epsfig{figure=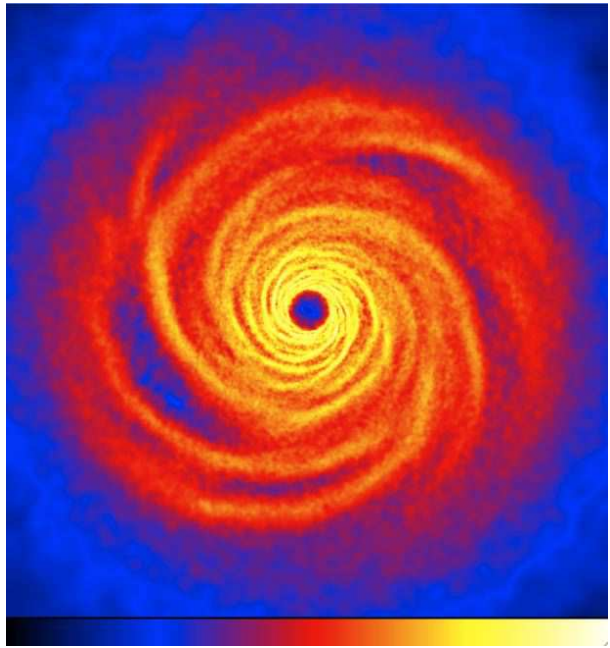,width=80mm}}
\caption{Surface density structure of a self-gravitating disc with
  $M_{\mathrm{disc}}= 0.25 M_{\odot}$ (the central star mass is
  $M_{\star}=1 M_{\odot}$). The spiral structure is a quasi-steady
  feature lasting for at least several thermal timescales. The panel
  shows the logarithm of the surface density $\Sigma$ with the scale
  covering $1<\log (\Sigma/\mathrm{g~cm}^{-2})<4.7$. The size of the
  box is 50 au across.}
\label{fig:gasspiral}
\end{figure} 

The simulations by \citet{LR04} indeed show the effectiveness of the
self-regulation process. The disc extends from $0.25$ au to $25$ au,
and is characterized initially by a surface density profile
$\Sigma\propto R^{-1}$ and a temperature profile $T\propto
R^{-1/2}$. The exact surface density is determined by specifying a
total disc mass, and the temperature is determined 
by specifying that the Toomre 
$Q$ parameter has an initial value of $2$ at the outer edge of the disc.
The temperature profile, however, is rapidly modified by
the competing heating and cooling processes operating in the disc, as
discussed above. At the end of the simulations a self-regulated state
is achieved with an almost constant profile of $Q$, with a value close
to unity. The spiral structure obtained in this way is a quasi-steady
feature lasting for at least several thermal timescales (i.e., at
least until the end of the simulations). Fig. \ref{fig:gasspiral}
shows the final disc structure for a disc whose total mass is
$M_{\mathrm{disc}}= 0.25M_{\odot}$. The image shows the logarithm of
$\Sigma$, with a colour scale covering $1<\log
(\Sigma/\mathrm{g~cm}^{-2})<4.7$.

The spiral structure transports angular momentum in the disc and
therefore promotes the accretion process. This leads also, in some
cases, to a steepening of the surface density profile. This process
however occurs on the much longer `viscous' timescale, so that the
final profile of $\Sigma$ is only slightly modified with respect to
the initial one (see \citealt{LR04}).

\subsection{Numerical setup}

We initially evolve the gas disc in the absence of any test particles,
as described in the previous Section. After $\sim 6$ orbital periods
at the outer edge of the disc (i.e., after $\sim 800$ yrs), a
quasi-steady, self-regulated state is reached. We then introduce the
planetesimals and evolve the simulation as described above in Section
\ref{sec:sph} for roughly one more outer orbital period (i.e., $\sim
125$ yrs).

\begin{figure*}
\centerline{\epsfig{figure=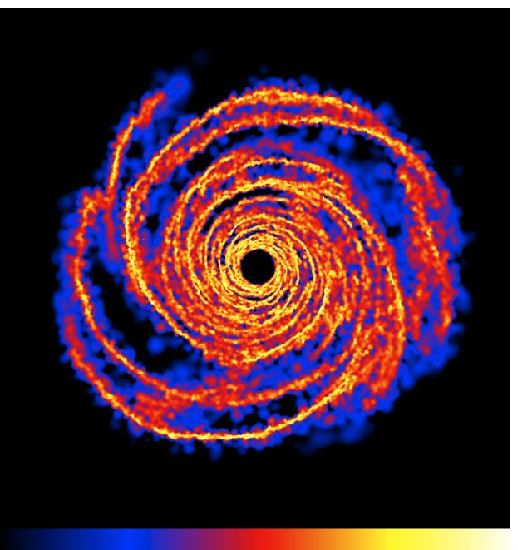,width=80mm}
            \epsfig{figure=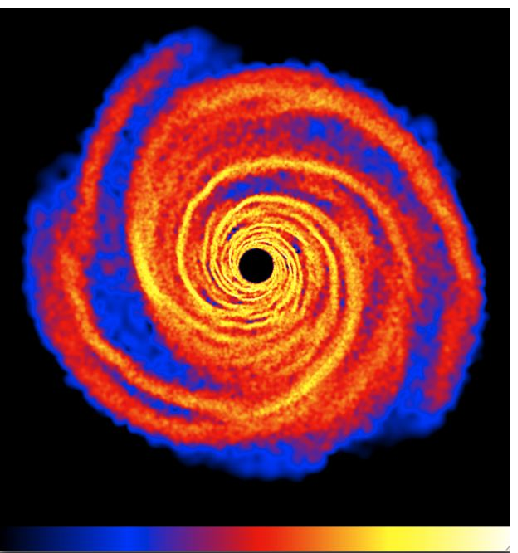,width=80mm}}
\centerline{\epsfig{figure=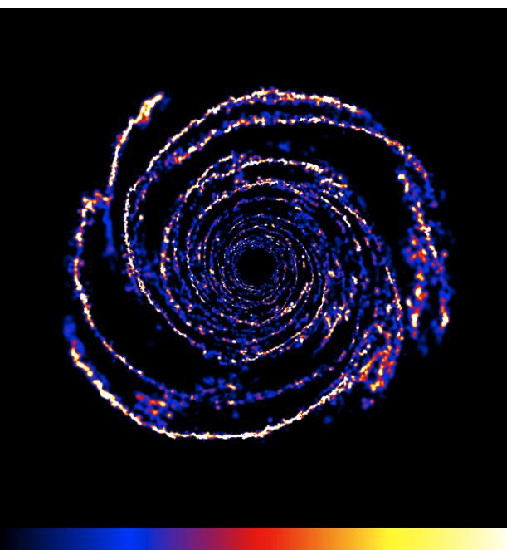,width=80mm}
            \epsfig{figure=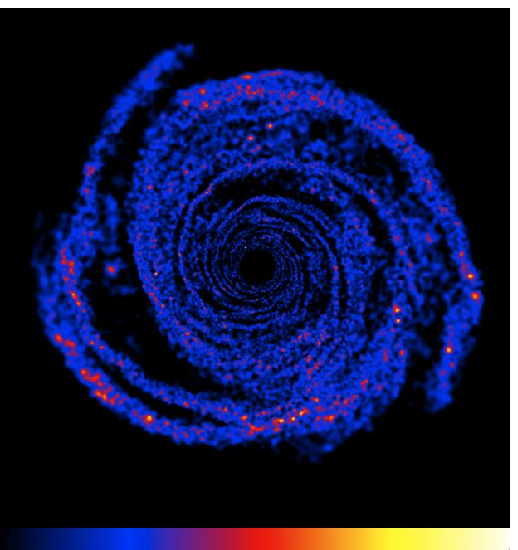,width=80mm}}
\caption{Surface density structure of the planetesimal discs after one
  outer orbital time. The upper panels show $\Sigma_{\mathrm{p}}$, for
  the 50 cm planetesimals (left panel) and for the 1000 cm
  planetesimals (right panel). The surface densities have been
  multiplied by $100$ and the colour scaling is the same as in
  Fig. \ref{fig:gasspiral}, in order to have a direct comparison with
  the gas surface density structure. The size of the box displayed is
  also the same as in Fig. \ref{fig:gasspiral}. The bottom panels show
  the ratio $\Sigma_{\mathrm{p}}/\Sigma$ for the two cases (the colour
  scales of these panel are the same in the two cases and cover
  $0.004< \Sigma_{\mathrm{p}}/ \Sigma<0.04$). If the gas and
  planetesimals respond in the same way to the gravitational
  instabilities, the density ratio would be uniform through the disc,
  so that any non-uniformity in the bottom panels is a direct measure
  of the concentration effect on the planetesimals caused by the
  combination of gas gravity and drag force. The planetesimal tend to
  concentrate in clumpy regions along the spiral arms, where the
  minima of the gravitational potential are located. The effect is
  particularly evident for planetesimal size of 50 cm, for which
  $\Sigma_{\mathrm{p}}/ \Sigma$ can be enhanced by more than a factor
  of 50.}
\label{fig:results}
\end{figure*} 

\begin{figure*}
\centerline{\psfig{figure=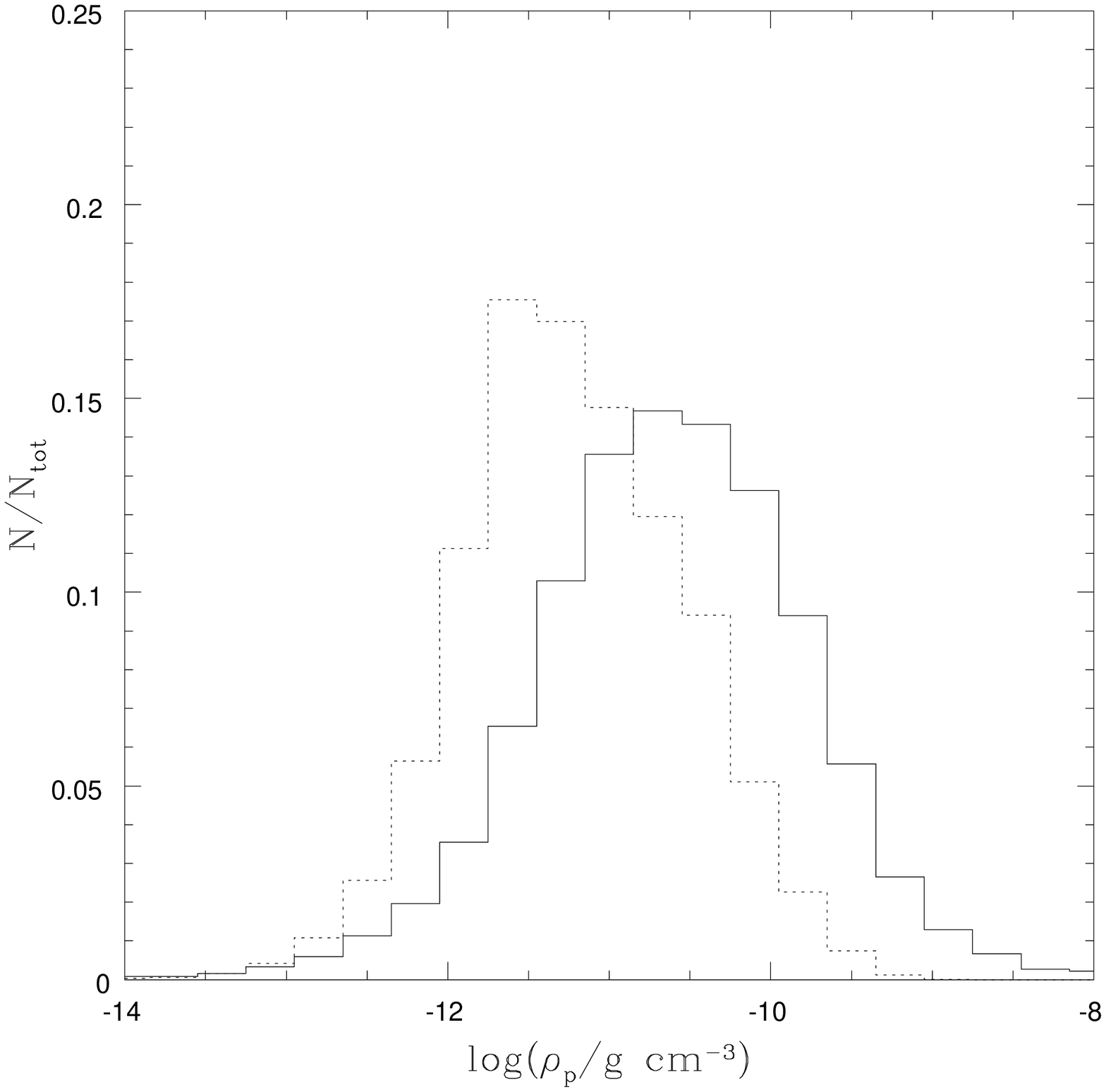,width=80mm}
            \psfig{figure=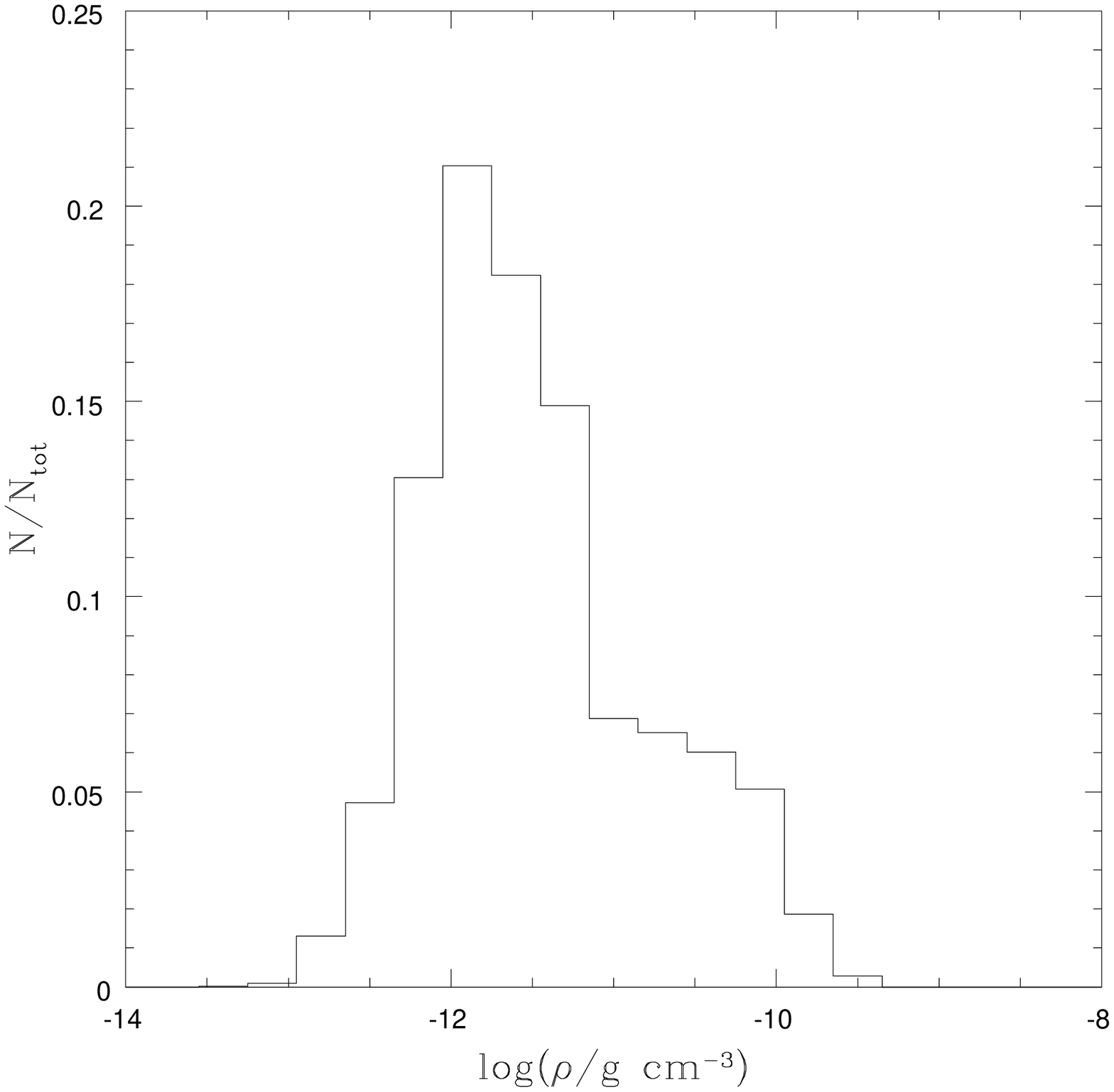,width=80mm}}
\caption{{\bf Left panel:} volume density distribution of the
  planetesimal for the two different sizes considered (solid line: 50
  cm; dotted line: 1000 cm), at the end of the simulation. Note that
  the particular shape of the density distribution is not very
  meaningful, since it depends on the initial surface density
  distribution that we assume. What is interesting is that, starting
  from the same initial density distribution, at the end of the
  simulation the 50 cm sized planetesimals can reach densities at
  least one order of magnitude larger than the 1000 cm case. {\bf
  Right panel:} corresponding gas density distribution at the end of
  the simulation, scaled down by two orders of magnitude for a direct
  comparison with the planetesimal density.}
\label{fig:hist}
\end{figure*} 

In order to cover a wide range of planetesimal sizes we have
considered separately planetesimal sizes of 50 cm and 1000 cm. Based
on the results of Section \ref{sec:dynamics}, we expect the 50 cm
sized planetesimal to have the largest radial drift. For smaller
planetesimal sizes (e.g, 1 cm) we expect the acceleration due to the
drag force to be very large, resulting in extremely long computation
times. However, for such small planetesimals, we expect the drag to be
so large that their structure will closely match that of the gas. In
addition to the 50 and the 1000 cm cases, we have also performed one
simulation in which no drag force was included, so as to provide a
direct measure of the effect of gas drag on the evolution of the
planetesimals.

The planetesimal disc initially extends from $R=2$ au to $R=20$
au. The surface density profile of the planetesimals
$\Sigma_{\mathrm{p}}$ was taken to be proportional to $R^{-1}$. Since
we neglect the planetesimals' self-gravity and the back reaction of
the drag force on the gas, the actual value of the planetesimal disc
surface density does not influence the results of the simulations (the
planetesimal SPH particles are just a ``tracer'' of the evolution of
the solid bodies in the gas disc). However, in order to present
illustration values in the analysis of our results, we will assume
that the initial ratio of the planetesimal to gas surface densities is
$0.01$ in all cases.

Initially all the planetesimals are located in the $z=0$
plane. However, during the simulation, the random motions induced by
the gravitational instabilities rapidly stir the planetesimal disc up,
so that eventually it acquires a finite thickness $H_{\mathrm{p}}$
slightly smaller than the gas disc thickness $H$.

\section{Results}
\label{sec:results}

Fig. \ref{fig:results} shows the surface density structure of the
planetesimal discs, one outer orbital time after the introduction of
the planetesimals in the simulations (i.e., after $\sim 125$ yrs). At
this stage, most of the disc has evolved for several dynamical
timescales, so that any initial transient features have
disappeared. The figure refers to the cases where the planetesimal
sizes were $50$ cm (left panels) and $1000$ cm (right panels). The
upper panels show the logarithm of the surface density
$\Sigma_{\mathrm{p}}$ (in order to have a direct comparison with
Fig. \ref{fig:gasspiral}, the surface densities have been multiplied
by $100$ and the same colour scale has been used). The bottom panels
show the ratio $\Sigma_{\mathrm{p}}/\Sigma$ of the planetesimal and
gas surface densities. The colour scales in the latter plots are
exactly the same for the two different planetesimal sizes and covers
the range $0.004< \Sigma_{\mathrm{p}}/ \Sigma<0.04$.

These plots clearly show how the planetesimal evolution changes with
changing planetesimal size. As expected, the 50 cm planetesimals are
strongly influenced by the gas drag and display a spiral pattern with
very thin spiral arms, indicating that the planetesimals are
concentrated at the bottom of the potential. The effect is reduced in
the 1000 cm case, where the spiral structure in the planetesimal disc
is similar to that of the gas disc, indicating that the planetesimals
are pushed into the spirals mainly because of the gravitational field
(note that since the planetesimals have no pressure support, we expect
the spiral arms to be slightly thinner even if no drag force is
introduced). A similar structure, with relatively broad spiral arms,
was indeed also seen in the simulation with no drag force.

The bottom panels in Fig. \ref{fig:results} show how the concentration
of planetesimals is modified by the combined effect of gas drag and
gravity. If the gas and planetesimals respond in the same way to the
gravitational instabilities, the density ratio would be uniform
through the disc, so that any non-uniformity in the bottom panels is a
direct measure of the concentration effect on the planetesimals caused
by the combination of gas gravity and drag force.  Clearly, the 50 cm
planetesimal reach a much higher concentration compared to the 1000 cm
case. At the end of the simulation $\Sigma_{\mathrm{p}}/\Sigma$ has
increased by a factor $\sim 3$ for the 1000 cm case, while in the 50
cm case the maximum increase can be as high as $\sim 50$, in some
regions therefore reaching surface densities comparable to that of the
gas.

\begin{figure}
\centerline{\epsfig{figure=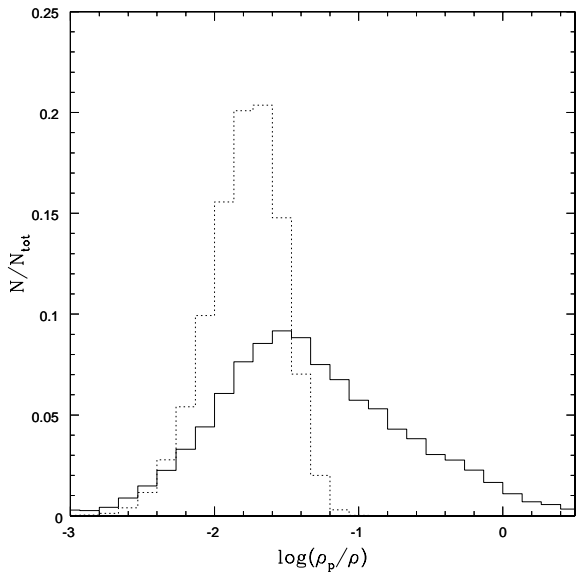,width=80mm}}
\caption{Distribution of volume density ratios (planetesimal/gas) for the two
different sizes considered (solid line: 50 cm; dotted line: 1000 cm)
at the end of the simulation. While the planetesimal/gas density ratio
is generally smaller than $0.05$ for the 1000 cm case (i.e., the
concentration enhancement is smaller than 5), in the 50 cm case it can
reach values of the order of unity (concentration enhancement $\sim
100$).}
\label{fig:ratio}
\end{figure}

Fig. \ref{fig:hist} shows a histogram of the distribution of
planetesimal volume densities, $\rho_{\mathrm{p}}$, for both the 50 cm
and 1000 cm particles (left panel), and a histogram of the
distribution of gas volume densities, $\rho$, scaled down by two
orders of magnitude for a direct comparison with the planetesimal
density (right panel), at the end of the simulations. In both panels
$N_{\mathrm{tot}}$ is the total number of particles of the type being
considered. The planetesimal volume densities were determined by
assuming, as mentioned earlier, that the initial planetesimal to gas
surface density ratio was $0.01$. The distribution for the 50 cm
planetesimals (solid line - left panel) shows a tail at high densities
extending to more than an order of magnitude above the 1000 cm
planetesimals (dotted line - left panel). The distribution of gas
volume densities (right panel) is also similar to that of the 1000 cm
particles, again illustrating that the 1000 cm particles are more
strongly influenced by the gravitational potential than by the drag
force.  The concentration of planetesimals is further illustrated in
Fig. \ref{fig:ratio} which shows the distribution of the ratio of the
planetesimal volume density to the gas volume density. This shows that
the volume density of the 50 cm planetesimals (solid line) may
increase to a value similar to the gas density.  The volume density
ratio for the 1000 cm particles (dotted line), on the other hand,
barely exceeds $0.1$ and in most regions has a value below
$0.05$. Given an ``unperturbed'' density ratio of 0.01, this means
that the volume density of the 1000 cm particles is almost never
enhanced by more than a factor of 10.  These results confirm our
expectations that for the planetesimals with the largest expected
radial drift, the combined effect of gas drag and spiral structure
induced by self-gravity leads to a significant concentration of the
planetesimals along the spiral arms.

\begin{figure*}
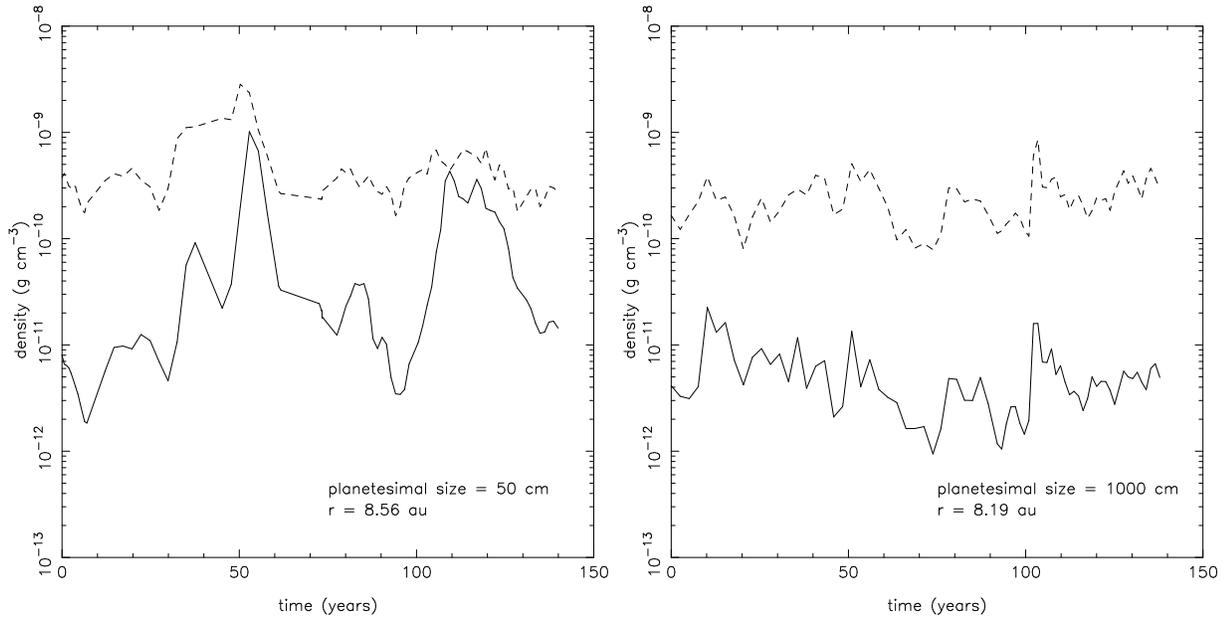

\centerline{\psfig{figure=fig6_1.ps,width=80mm}
            \psfig{figure=fig6_2.ps,width=80mm}}
\caption{Time evolution of the densities for a 50 cm planetesimal
  (left panel) and for a 1000 cm planetesimal (right panel). The solid
  line shows the planetesimal density, while the dashed line shows the
  gas density.}
\label{fig:time}
\end{figure*} 

Fig. \ref{fig:time} shows how the volume density, as seen by a
representative planetesimal located at $\sim 8$ au, varies with time.
The left and the right panels display the results for the 50 cm and
the 1000 cm cases, respectively. The solid line shows the planetesimal
volume density, while the dotted line shows the corresponding gas
volume density. The differences between the two cases are striking
also in this case. The planetesimal density for the 1000 cm particles
essentially follows closely the gas density, with relatively small
variations, and oscillates between high and low values as the
planetesimal goes in and out of the regions of enhanced gas density
(the spiral arms). In the 50 cm case, the planetesimal density reaches
extremely high values becoming comparable to the gas density when the
planetesimal moves into a spiral arm. The planetesimal density can
also remain high for as long as $\sim 20$ yrs, comparable to the
dynamical timescale at 8 au.

\begin{figure}
\centerline{\epsfig{figure=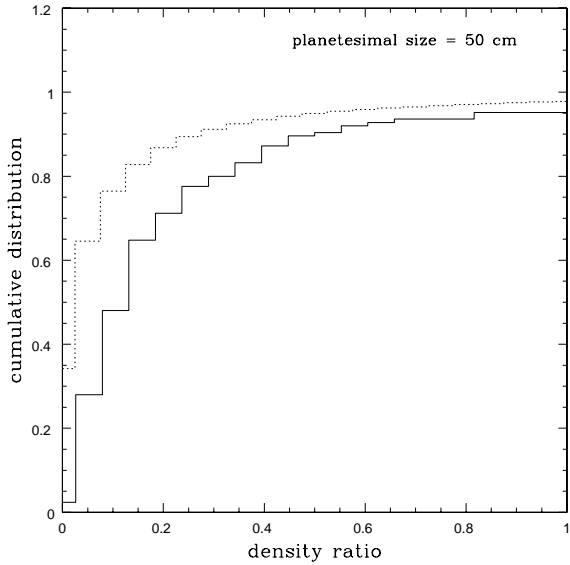,width=80mm}}
\caption{{\bf Solid line}: cumulative distribution of the maximum of
  the density ratio $\rho_{\mathrm{p}}/\rho$ attained during the run.
  {\bf Dotted line}: cumulative distribution of
  $\rho_{\mathrm{p}}/\rho$ at a given time. Both plots refer to the 50
  cm case. While at any given time during the simulation the fraction
  of planetesimals with $\rho_{\mathrm{p}}/\rho>0.1$ (i.e.,
  $\rho_{\mathrm{p}}/\rho$ enhanced by more than a factor 10) is
  $\sim 25$\%, the fraction of planetesimal that at some stage
  during the simulation attains the same value of
  $\rho_{\mathrm{p}}/\rho$ is significantly higher, $\sim 75$\%.}
\label{fig:cumulative}
\end{figure}

To summarize, Fig. \ref{fig:ratio} shows that at a given time a
significant fraction of the 50 cm planetesimals have a large
concentration (say, $\rho_{\mathrm{p}} /\rho>0.1$). On the other hand,
Fig. \ref{fig:time} shows that a given particle spends only a fraction
of the total simulation at high $\rho_{\mathrm{p}}/ \rho$. This
suggest that the total fraction of planetesimals which at some stage
during the run have a large $\rho_{\mathrm{p}}/\rho$ is actually
larger than the corresponding fraction taken at a given time. To
estimate this effect we have computed the maximum value of
$\rho_{\mathrm{p}}/\rho$ attained by the planetesimals during the
whole simulation (in order to reduce computational time, we have
performed this analysis only for a subset of the total number of
planetesimal SPH particles). The dotted line in 
Fig. \ref{fig:cumulative} shows 
the cumulative distribution of $\rho_{\mathrm{p}}/\rho$
at a given time (i.e., the cumulative distribution corresponding to
the solid line of Fig. \ref{fig:ratio}), while the solid line shows
the cumulative distribution of the maximum $\rho_{\mathrm{p}}/\rho$, 
computed as described above. 
This figure clearly shows that, while at any given
time during the simulation the fraction of planetesimals with
$\rho_{\mathrm{p}}/\rho>0.1$ (i.e., $\rho_{\mathrm{p}}/\rho$ enhanced
by more than a factor 10) is $\sim 25$\%, the fraction of planetesimal
that at some stage during the simulation attains the same value of
$\rho_{\mathrm{p}}/\rho$ is significantly higher, $\sim 75$\%.

\section{Discussion and conclusions}

The concentration of planetesimal, due to the combined effect of gas
drag and gravity, that we find in our simulations can have a significant
effect on the process of coagulation of planetesimal into larger
bodies. This can occur either by increasing the planetesimal collision
rate and/or by making the planetesimal sub-disc gravitationally
unstable.

\begin{figure}
\centerline{\epsfig{figure=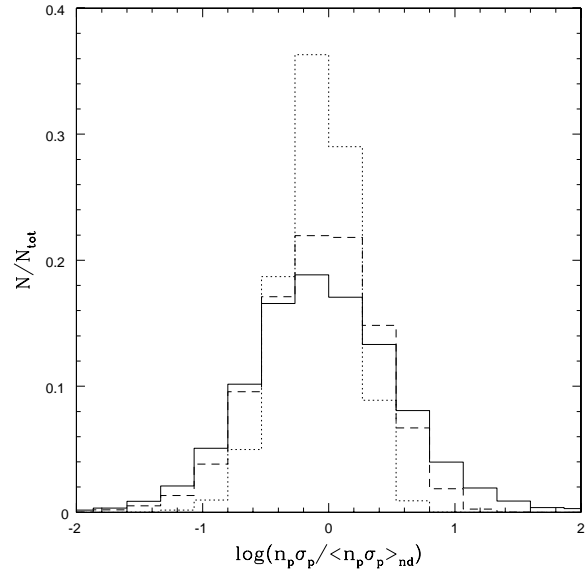,width=80mm}}
\caption{Distribution of $n_{\mathrm{p}} \sigma_{\mathrm{p}} /\langle
  n_{\mathrm{p}} \sigma_{\mathrm{p}} \rangle_{\mathrm{nd}}$ for the 50
  cm case (solid line), for the 1000 cm case (dashed line) and for the
  no drag simulation (dotted line). Fluctuations of $n_{\mathrm{p}}
  \sigma_{\mathrm{p}} $ not related to the gas drag result only in an
  increase of the collision rate by no more than a factor $\sim
  6$. In contrast, the introduction of the gas drag leads to a broader
  distribution of collision rates, especially for the 50 cm case,
  where the distribution has a tail extending up to two orders of
  magnitude above the average value.}
\label{fig:collision}
\end{figure} 

The collision rate of planetesimals is proportional to
$n_{\mathrm{p}}\sigma_{\mathrm{p}}$, where
$n_{\mathrm{p}}=\rho_{\mathrm{p}}/m_{\mathrm{p}}$ is the planetesimal
number density and $\sigma_{\mathrm{p}}$ is their velocity
dispersion. Since in our case the Safronov number
$\Theta=Gm_{\mathrm{p}}/2\sigma_{\mathrm{p}}^2a$ is always much
smaller than unity, we can neglect the effect of gravitational
focusing. In order to assess the effect of gas drag on the collision
rate, we have first computed the azimuthally averaged value of
$n_{\mathrm{p}}\sigma_{\mathrm{p}}$ from the simulation with no drag
force, $\langle n_{\mathrm{p}}\sigma_{\mathrm{p}}
\rangle_{\mathrm{nd}}$, as a function of radius. We have then
computed, for every planetesimal SPH particle in both the 50 cm case
and the 1000 cm case, the ratio $n_{\mathrm{p}} \sigma_{\mathrm{p}}
/\langle n_{\mathrm{p}} \sigma_{\mathrm{p}} \rangle_{\mathrm{nd}}$,
where the average value is computed at the same radial location in the
disc. If the gas drag had no effect on the collision rate, the
distribution of $n_{\mathrm{p}} \sigma_{\mathrm{p}} /\langle
n_{\mathrm{p}} \sigma_{\mathrm{p}} \rangle_{\mathrm{nd}}$ would be
strongly peaked around unity. Fig. \ref{fig:collision} shows the
distribution of $n_{\mathrm{p}} \sigma_{\mathrm{p}} /\langle
n_{\mathrm{p}} \sigma_{\mathrm{p}} \rangle_{\mathrm{nd}}$ that we have
obtained in the three cases (no drag force: dashed line; 1000 cm size:
dotted line; 50 cm size: solid line). As expected, the distribution
for the no-drag simulation is strongly peaked around
unity. Fluctuations of $n_{\mathrm{p}} \sigma_{\mathrm{p}}$ not
related to the gas drag result only in an increase of the collision
rate by no more than a factor $\sim 6$. In contrast, the introduction
of the gas drag leads to a broader distribution of collision rates,
especially for the 50 cm case where the distribution has a tail
extending more than two orders of magnitude above the average
value. As discussed earlier, the fraction of particles concentrated in
the spiral arms at a given time is smaller than the fraction of
particles that, during the course of the whole simulation, are at some
stage concentrated in the spiral arms. Since the enhancement in
collision rate is due to the enhanced density resulting from the
concentration of the planetesimals in the spiral arms, the number of
particles over the entire simulation time that at some stage are in a
region of enhanced collision rate will also be greater than the number
at a single time. Depending on how well particles of this size stick
together during collisions, this enhanced collision rate could play an
important role in the growth of larger planetesimals.

\begin{figure*}
\centerline{\psfig{figure=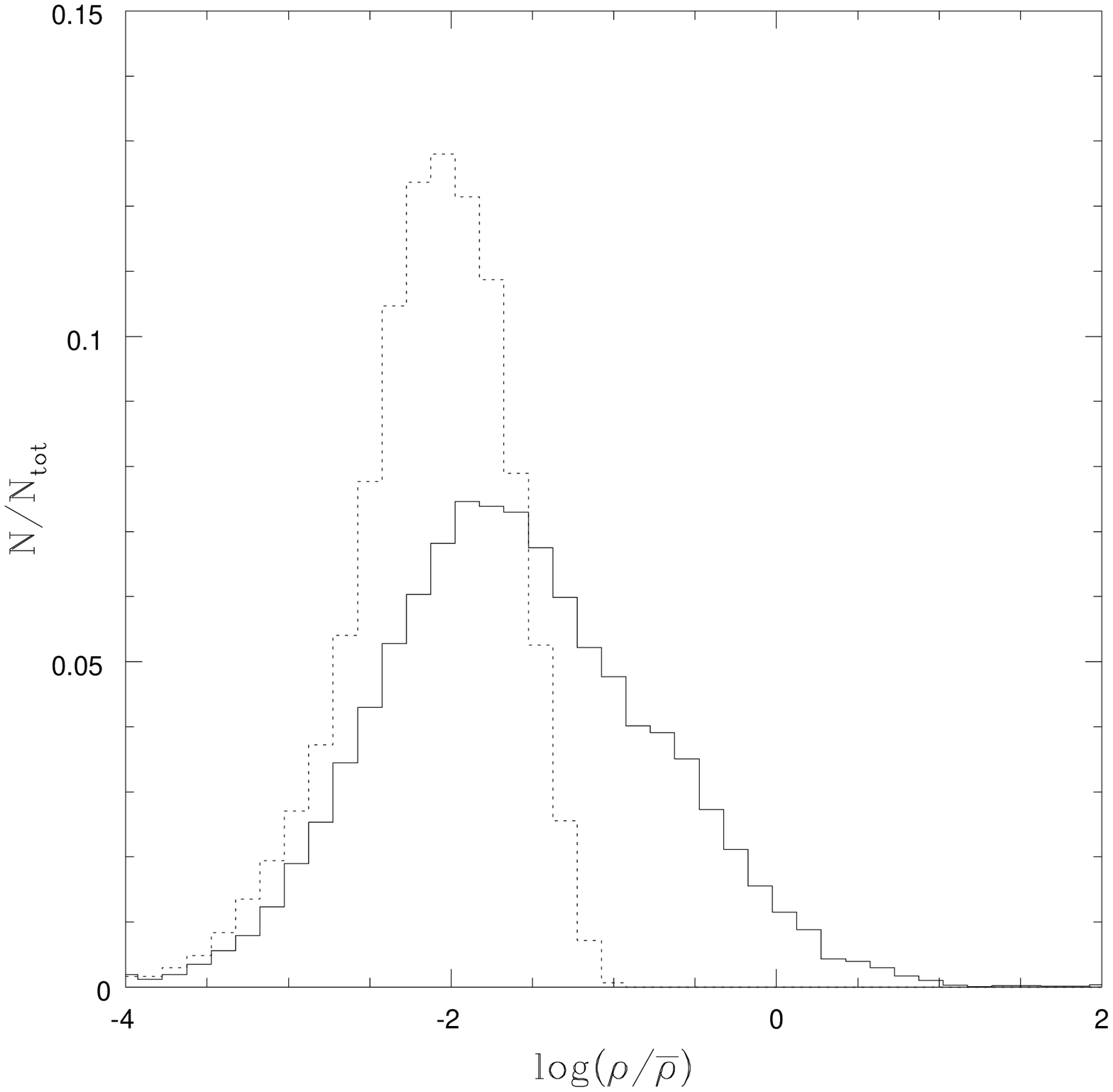,width=80mm}
            \psfig{figure=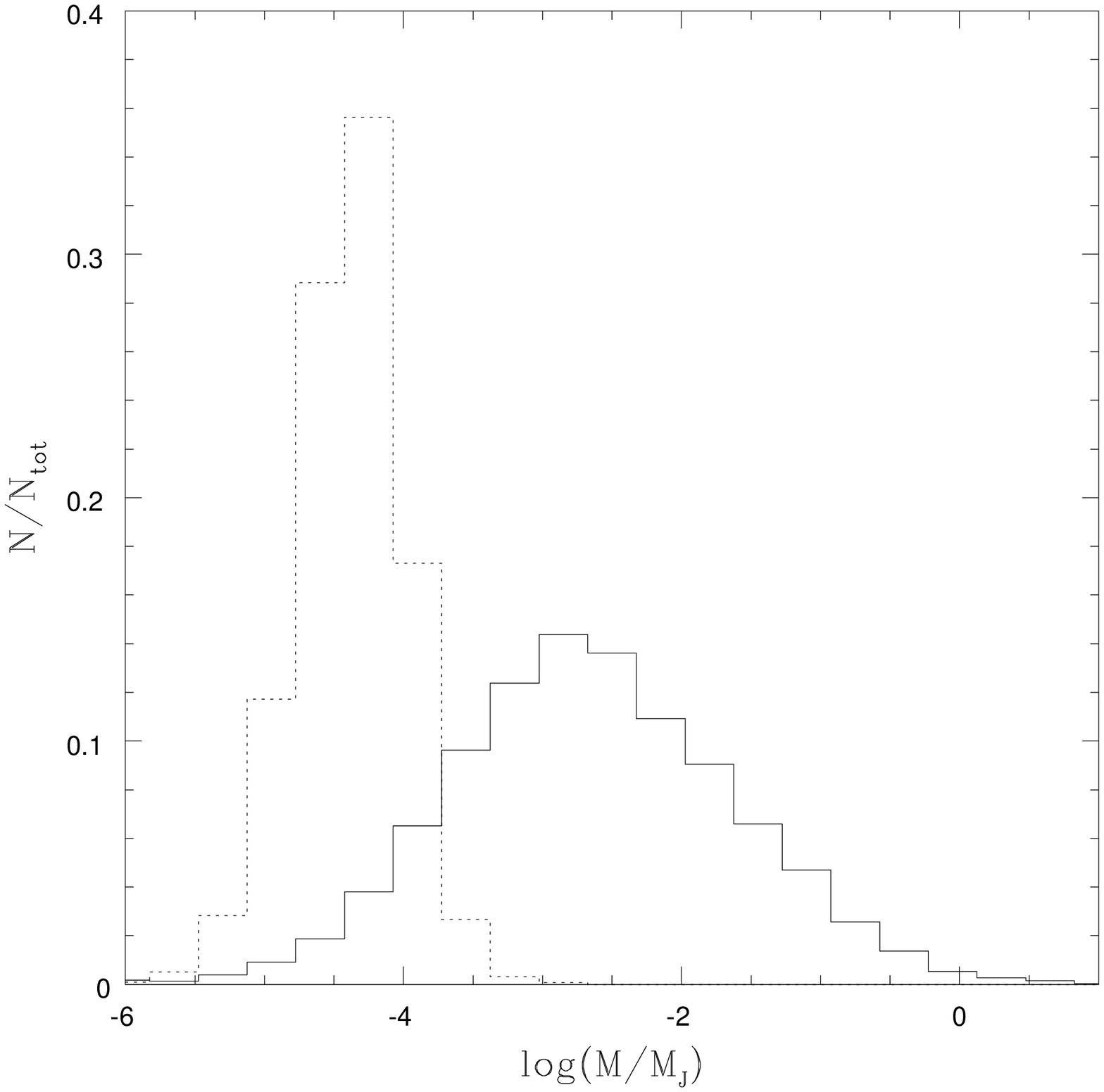,width=80mm}}
\caption{Left panel: distribution of $\rho_{\mathrm{p}}/\bar{\rho}$
  for the 50 cm (solid line) and the 1000 cm case (dotted line). Right
  panel: distribution of $M/M_{\mathrm{J}}$ in the two cases. These
  results show that some of the high density clumps observed in the 50
  cm case should be subject to gravitational instability.}
\label{fig:gravinst}
\end{figure*} 

A planetesimal surface density enhancement of a factor $\sim 20$
may also be sufficient to make the planetesimal sub-disc gravitationally
unstable (e.g., \citealt{youdinshu02}). This is exactly the range of
concentrations that we achieve in 50 cm simulation. However, since we 
have neglected the planetesimal self-gravity, we are not able to 
obtain a gravitational instability in the planetesimal disc in
our simulations (for a more detailed study of the stability of a two
component self-gravitating disc, see \citet{bertin88}).
In order to assess the importance of the planetesimal
self-gravity we have performed two separate tests. 

We have first computed the quantity
\begin{equation}
\frac{\rho_{\mathrm{p}}}{\bar{\rho}}=\frac{\rho_{\mathrm{p}}
  R^3}{M_{\star}}=\frac{G\rho_{\mathrm{p}}}{\Omega_{\mathrm{K}}^2},
\end{equation}
where $R$ is the radius of a region that has a local planetesimal
density of $\rho_{\mathrm{p}}$. This quantity is a measure of the
relative effects of local gravitational collapse for the planetesimals
versus tidal disruption.  Fig. \ref{fig:gravinst} shows the distribution
of $\rho_{\mathrm{p}}/ \bar{\rho}$ in the 50 cm case (solid line) and in
the 1000 cm case (dotted line). In the 50 cm case $\rho_{\mathrm{p}}/
\bar{\rho}$ reaches much higher values than in the 1000 cm case,
becoming as large as $\sim$ 1. This suggest that the density
enhancements in the 50 cm simulation may be gravitationally
significant and that the planetesimal disc could become
gravitationally unstable.

As a separate measure, we have also computed the Jeans mass,
\begin{equation}
M_{\mathrm{J}}=\frac{1}{6}\pi\rho_{\mathrm{p}}\left
(\frac{\pi\sigma_{\mathrm{p}}^2}{G\rho_{\mathrm{p}}} \right)^{3/2},
\end{equation}
at the location of every planetesimal particle. If, at any location,
the minimum resolvable mass (the mass, $M$, contained within 2
smoothing lengths) exceeds the local Jeans mass, then the
planetesimals' self-gravity has to play a role. The distribution of
$M/M_{\mathrm{J}}$ is shown in the right panel of Fig.
\ref{fig:gravinst}. In the 50 cm case, the distribution is shifted
towards higher values and is much broader than in the 1000 cm
case. For the most massive clumps, the ratio $M/M_{\mathrm{J}}$ can
become comparable to or even larger than unity, indicating that the
gravitational instability would play a significant role in these regions.

Note that since the vertical scale height of the planetesimal disc is
comparable to the scaleheight of the gas disc, the high densities
achieved in our simulation using 50 cm particles is only due to radial
and azimuthal compressions rather than by the vertical settling of the
planetesimals in the midplane. Therefore, unlike in the standard
picture for gravitational instability
\citep{goldreichward,youdinshu02}, our results should not be affected
by additional turbulence generated by the vertical shear between the
gas and the dust.

To summarize, in this work we have shown how the interaction between a
self-gravitating gaseous protoplanetary disc and embedded
planetesimals plays an important role in accelerating planetesimal
growth. However, there are a number of important effects that we have
neglected in this first approach to the problem. {\it (i)} The
planetesimals have been essentially modelled as test particles,
ignoring both the back reaction of the planetesimals on the gas and
the planetesimal self-gravity. In regions where the planetesimal
density is enhanced (becoming in some regions comparable to the gas
density) both effects may be very important. {\it (ii)} In each
simulation the planetesimals are assumed to be of a single size and
the volume density is computed by assuming that the initial surface
density ratio is $0.01$. In reality there will be a range of
planetesimal sizes (e.g., \citealt{mathis77,mizuno88}; note, however,
that these studies consider grain sizes much smaller than those
considered here) and only those with sizes, in this case, between
$\sim 10$ and $100$ cm (see Fig. \ref{fig:drift}) will be
significantly influenced by the self-gravitating structures in the gas
disc. During the evolution of the protoplanetary nebula there might
well be some stage where most of the planetesimal mass is contained
within a relatively small range of sizes, so that our assumptions may
not be unrealistic. When this size range includes the size for which
the drift induced by the drag is significant, we can expect a
significant increase in collision rate and an enhanced tendency toward
gravitational collapse. To address the details of these processes we
would need to consider many other effects (such as the sticking
properties of planetesimals) which are beyond the scope of the present
paper. {\it (iii)} In this work we have considered, as an
illustration, the self-gravitating structure resulting from a
relatively massive gas disc (with a mass $\sim
0.25M_{\star}$). However, a well defined spiral structure is also
present in significantly less massive discs \citep{LR04}. It would 
then be interesting to check the dependence of some of the details of the
results described here (such as what is the relevant size range for
the planetesimals that display the strongest response to the spiral
structure) on the specific choice of the gas disc properties.

We plan in the future to include some of the effects described above,
such as the backreaction of the planetesimals on the disc gas, the
effect of planetesimal self-gravity, and we plan to consider various
disc masses. However, it seems clear that if protoplanetary discs
experience a self-gravitating phase, the resulting disc structures
could well play an important role in planetesimal evolution and growth
and could ultimately influence the growth of terrestrial planets and
the cores of gas/ice giant planets. Also, since a protoplanetary disc
is most likely to become gravitationally unstable early in the star formation
process, we might expect substantial processing of the dust prior to the
optically visible Classical T Tauri phase.

\label{sec:conclusion}

\section*{Acknowledgements}

The simulations reported in this paper made use of the UK
Astrophysical Fluids Facility (UKAFF). WKMR acknowledges support from
a UKAFF Fellowship. GL acknowledges support from the EU Research Training Network {\em Young Stellar Clusters}. JEP acknowledges support from the STScI visitor
program. PJA acknowledges support from the National Science Foundation
under grant AST 0407040, and from NASA under grant NAG5-13207 
issued through the Office of Space Science. We thank Cathie Clarke 
and Steve Lubow for interesting discussions.

\bibliographystyle{mn2e}
\bibliography{lodato}

\end{document}